\documentclass[aip,graphicx]{revtex4-1}
\usepackage{graphicx} 
\usepackage{xcolor}
\usepackage{float} 
\draft 

\begin{document}


\title{185\,mW, 1\,MHz, 15\,fs carrier-envelope phase-stable pulse generation via polarization-optimized down-conversion from gas-filled hollow-core fiber} 


\author{Anchit Srivastava}
\affiliation{Max Planck Institute for the Science of Light, Staudstrasse 2, Erlangen, 91058, Germany.}%
\affiliation{Friedrich-Alexander-Universit{\"a}t Erlangen-N{\"u}rnberg, Staudstrasse 7, 91058 Erlangen, Germany.}%

\author{Kilian Scheffter}%
\affiliation{Max Planck Institute for the Science of Light, Staudstrasse 2, Erlangen, 91058, Germany.}%
\affiliation{Friedrich-Alexander-Universit{\"a}t Erlangen-N{\"u}rnberg, Staudstrasse 7, 91058 Erlangen, Germany.}%

\author{Soyeon Jun}%
\affiliation{Max Planck Institute for the Science of Light, Staudstrasse 2, Erlangen, 91058, Germany.}%
\affiliation{Friedrich-Alexander-Universit{\"a}t Erlangen-N{\"u}rnberg, Staudstrasse 7, 91058 Erlangen, Germany.}%

\author{Andreas Herbst}%
\affiliation{Max Planck Institute for the Science of Light, Staudstrasse 2, Erlangen, 91058, Germany.}%
\affiliation{Friedrich-Alexander-Universit{\"a}t Erlangen-N{\"u}rnberg, Staudstrasse 7, 91058 Erlangen, Germany.}%

\author{Hanieh Fattahi}%
 \email{hanieh.fattahi@mpl.mpg.de.}
\affiliation{Max Planck Institute for the Science of Light, Staudstrasse 2, Erlangen, 91058, Germany.}%
\affiliation{Friedrich-Alexander-Universit{\"a}t Erlangen-N{\"u}rnberg, Staudstrasse 7, 91058 Erlangen, Germany.}%


\date{\today}

\begin{abstract}
Gas-filled hollow core fibers allow the generation of single-cycle pulses at megahertz repetition rates. When coupled with difference frequency generation, they can be an ideal driver for the generation of carrier-envelope phase stable, octave-spanning pulses in the short-wavelength infrared. In this work, we investigate the dependence of the polarization state in gas-filled hollow-core fibers on the subsequent difference frequency generation stage. We show that by adjusting the input polarization state of light in geometrically symmetric systems, such as hollow-core fibers, one can achieve precise control over the polarization state of the output pulses. Importantly, this manipulation preserves the temporal characteristics of the ultrashort pulses generated, especially when operating near the single-cycle regime. We leverage this property to boost the down-conversion efficiency of these pulses in a type I difference frequency generation stage. Our technique overcomes the bandwidth and dispersion constraints of the previous methods that rely on broadband waveplates or adjustment of crystal axes relative to the laboratory frame. This advancement is crucial for experiments demanding pure polarization states in the eigenmodes of the laboratory frame.
\end{abstract}


\maketitle 

\section{\label{sec:level1}Introduction}

Carrier-envelope phase stable (CEP) broadband sources \cite{baltuvska2002controlling} in the short-wavelength infrared (SWIR) and mid-infrared (MIR) ranges are crucial for many applications where the interaction between matter and light is sensitive to single oscillations of light, such as in strong field physics \cite{vozzi2012strong,schultz2007strong, ghimire2019high,chevreuil2021water, fattahixray, app8050728} or field-resolved metrology \cite{PhysRevB.107.054302, Hogue:23, herbst2022recent, yeung2023lightwave, zimin2023electricfieldresolved, Kempf2023-bf, Mamaikin:22, srivastava2023near}. It has been shown that CEP stable pulses can be generated by actively stabilizing optical oscillators or fiber lasers \cite{Vasilyev:19, Kowalczyk:23, Schoenfeld:22}, or by passive stabilization techniques such as frequency down-conversion \cite{CEPstabCerullo, Reiger:24, ernotte2016frequency}. Since oscillators generate pulses with low peak power, optical parametric amplification has been utilized to enhance the energy of the ultrashort pulses in these spectral regions \cite{ Elu:17, Deng:12, Alismail:17, ishii2011carrier, Fattahi:14}. However, parametric amplifiers require high-energy pump sources, which mostly operate at kilohertz repetition rates. The amplified pulses often exhibit additional phase jitter in their CEP, stemming from the need for interferometric stability between the pump and seed pulses \cite{Schwarz:12, Fattahi:12}. Passive CEP stabilization based on intrapulse difference frequency generation (IPDFG) offers unparalleled stability and minimal CEP fluctuations, making it the preferred method for generating CEP-stable, octave-spanning pulses at megahertz repetition rates. 

In IPDFG, down-converted passively CEP-stable pulses are generated through nonlinear frequency mixing of spectral components within a broadband pump pulse in a $\chi^{(2)}$ nonlinear crystal. Two types of phase-matching can be used to compensate for the phase velocity mismatch between the interacting beams. In type II phase-matching, the input pump and IPDFG pulses have orthogonal polarizations. On the contrary, in type I phase-matching, the input pump pulses are projected into both orthogonal axes of the nonlinear crystal. This projection compensates for the phase velocity mismatch between the high-frequency and low-frequency components of the pump, as well as the IPDFG pulses. Consequently, enhancing the conversion efficiency in type I phase-matching requires optimizing the projection of the pump's spectral components onto the crystal's orthogonal axes \cite{Fattahi:13}. To meet this requirement, the optical axis of the nonlinear crystal can be rotated such that the crystal axes have an angle with respect to the linear polarization of the pump beam in the laboratory frame. In this case, the polarization of the newly generated IPDFG pulses has the same angle relative to the linearly polarized pulse in the laboratory frame, which poses a challenge for polarization-sensitive applications such as electro-optic sampling, spectroscopic ellipsometry, polarization spectroscopy, or optical parametric amplification \cite{Kopf:21, ASPNES2014334, Keiber2016-xv, doi:10.1126/sciadv.aax3408, doi:10.1021/acsphotonics.3c01322, 8742571, Wang:17}. 
Broadband half-wave plates present an alternative option for optimizing the spectral distribution of input pump pulses along the crystal axes. However, their effectiveness with ultrashort pulses is constrained due to non-uniform phase retardation across a broad bandwidth and additional material dispersion. These constraints lead to temporal broadening of input pump pulses and lower conversion efficiency.

\begin{figure}[h]
\begin{center}
  \includegraphics[width=1\linewidth]{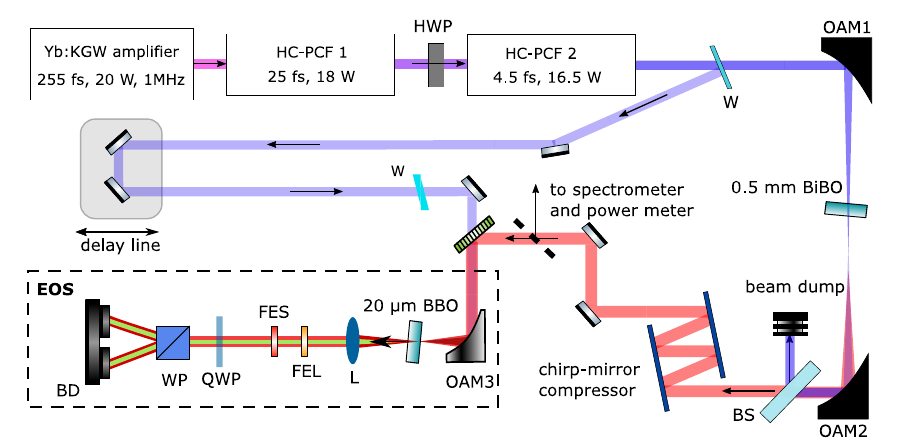}
  \caption{\textbf{Experimental setup.} The experimental setup for IPDFG comprises a two-stage gas-filled hollow-core fiber setup pumped by a commercial Yb:KGW amplifier at 1.03\,$\mu$m. A half-wave plate is placed before the second fiber stage to tune the input polarization. The near-single cycle pulses in the fiber were down-converted in a 0.5\,mm-thick BiBO crystal. Electro-optic sampling was employed for the complete characterization of the generated pulses. HC:PCF, hollow-core photonic crystal fiber; HWP, half-wave plate; OAM, 90$^{\circ}$ off-axis parabolic mirror; BS, beam splitter; W, wedge pair; EOS, electro-optic sampling; L, lens; FEL, long-pass filter; FES, short-pass filter; QWP, quarter-wave plate; WP, Wollaston prism; BD, balanced photodiode.}
  \label{fig:fig1}
  \end{center}
\end{figure}

Gas-filled hollow-core fibers (HCF) are widely utilized to create few-cycle ultrashort pulses at high peak power and megahertz repetition rates with a high polarization purity \cite{russell2014hollow,taranta2020exceptional}. In capillaries or single-ring hollow-core fibers with an M-fold symmetry, where M indicates the number of tiny capillaries, the intrinsic birefringence is negligible. Therefore, it is anticipated that the nonlinear propagation dynamics within the fiber remain unaffected relative to the polarization state of the input pulses \cite{agrawal2000nonlinear}. In this study, we leverage the inherent symmetry of HCF to address the challenge of optimizing the spectral intensity distribution in type I IPDFG. By utilizing a waveplate to adjust the polarization state of the narrowband input pulses directed into a gas-filled HCF, we fine-tune the polarization state of near-single-cycle ultrashort pulses for efficient frequency down-conversion in subsequent stages. Our findings indicate that the effect of polarization rotation on the nonlinear dynamics within the fiber is minimal and does not alter the temporal duration of the generated near-single-cycle pulse. We demonstrate that enhancing the down-conversion efficiency can be achieved by optimizing the polarization rotation of the input pump directed to the fiber. The field-resolved measurement of the IPDFG pulses confirms that this optimization does not affect the electric field of the generated IPDFG pulses. 

\section{\label{sec:level2}Results}
The schematic of the setup used to generate and characterize IPDFG pulses is shown in figure \ref{fig:fig1} and presented with details in \cite{srivastava2023near}. The frontend was driven by a 1030\,nm Yb:KGW amplifier with an average power of 20\,W at 1\,MHz repetition rate and a pulse duration of 255\,fs. Two nonlinear gas-filled photonic crystal fiber stages were used to generate near-single cycle pulses. The spectral broadening in the first fiber stage was based on self-phase modulation, followed by dispersive mirrors to temporally compress the pulses to 25\,fs at the full-width half maximum (FWHM). In the second fiber stage, soliton-effect self-compression was used to compress the pulses further to a pulse duration of 4.5\,fs at the FWHM. The system ultimately delivered 16.5\,W of average power with an overall efficiency of 82\% and polarization extinction ratio of 98\%. 

\subsection*{Numerical analysis}

Bismuth borate (BiBO) crystal was chosen for frequency down-conversion due to its wide transmission range from 286\,nm to 2500\,nm, high damage threshold, and large effective nonlinear coefficient. BiBO exhibits a higher nonlinear coefficient compared to other suitable crystals at this range, such as lithium triborate (LBO), beta barium borate (BBO), and potassium deuterium phosphate (KDP) \cite{Ghotbi:04}. Numerical simulations were conducted in a simulation system for optical systems code (SISYFOS) \cite{arisholm1997general} to gain further insight into the phase-matching process. Both types of phase-matching were considered for down-conversion of the octave-spanning spectrum in a 0.5\, mm-thick BiBO crystal. For the type I phase-matching, the input polarization was split between the crystal's orthogonal axes. In this paper, we refer to these axes as `o' and `e' to represent ordinary and extraordinary crystal axes, respectively.

\begin{figure}[h]
\begin{center}
  \includegraphics[width=0.9\linewidth]{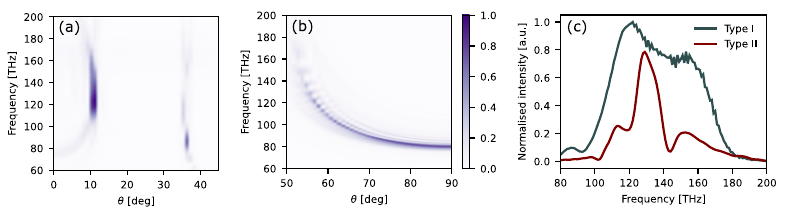}
  \caption{\textbf{Phase-matching simulation for 0.5\,mm-thick BiBO.} (a) IPDFG spectrum at various phase-matching angles for the (a) type I and (b) type II phase-matching, respectively. The broadest bandwidth for phase-matching is at $\theta$ = 11\,$^o$, whereas for type II is at $\theta$ = 55\,$^o$.}
  \label{fig:fig2}
  \end{center}
\end{figure}

We assumed a polarization ratio of 95\%-5\% (o-e) for simulating the type I phase-matching. The temporal profile of the output pulses from the frontent was characterized by second-harmonic frequency-resolved optical gating (SH-FROG) \cite{kane1993characterization}. The retrieved spectrum and retrieved temporal profile of the SH-FROG characterization were used as pump pulses for the IPDFG simulation. Figure \ref{fig:fig2}-a and figure \ref{fig:fig2}-b compare the spectral bandwidth of the IPDFG pulses versus the phase-matching angle of $\theta$ for type I and type II phase-matching. It is seen that the broadest spectral bandwidth is achieved at $\theta$ = 11\,$^o$ for type I phase-matching and at $\theta$ = 55\,$^o$ for type II phase-matching. The spectra of both phase-matching types at the optimized angles are shown in figure \ref{fig:fig2}-c. Both spectra are normalized to the energy of the IPDFG pulses in each case, indicating a broader spectral bandwidth and higher gain in type I phase-matching compared to type II.

\begin{figure}[h]
\begin{center}
  \includegraphics[width=1\linewidth]{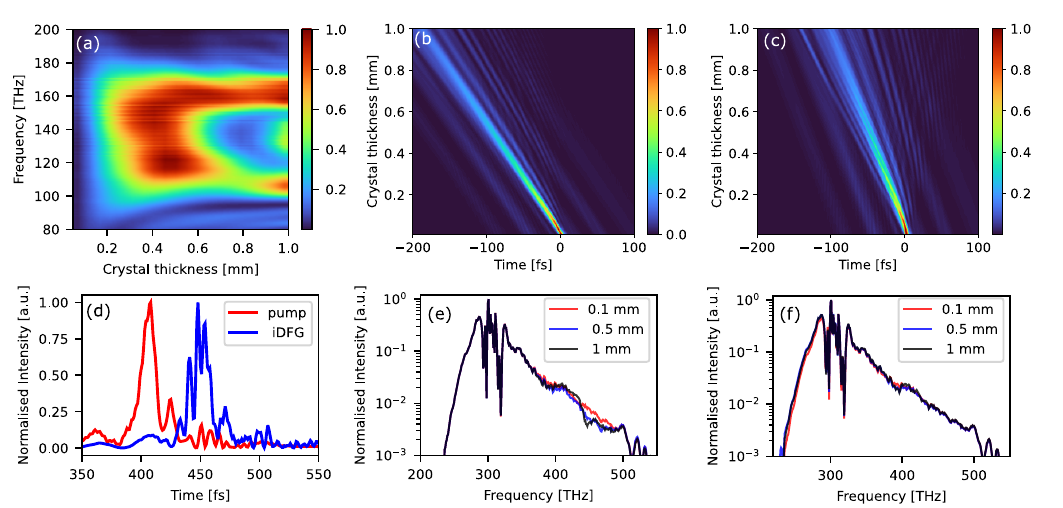}
  \caption{\textbf{Numerical simulation.} (a) Spectral evolution of the generated pulses over a 1\,mm-thick, type I, BiBO crystal at $\theta$ = 11\,$^o$. Numerically simulated temporal evolution of (b) pump pulse propagating and (b) IPDFG pulses through a 0.5\,mm BiBO crystal. (d) Temporal walk-off between the pump and IPDFG pulse after propagation in 0.5\,mm of BiBO crystal. Comparison of the output (e) o-pump and (f) e-pump at three different crystal thicknesses.}
  \label{fig:fig3}
  \end{center}
\end{figure} 

A one-dimensional simulation was performed to study the efficiency and bandwidth scaling of the IPDFG pulses in type I phase-matching versus crystal thicknesses. As shown in figure \ref{fig:fig3}-a, for a crystal thickness longer than 0.5\,mm, the center of mass of the difference frequency spectrum shifts to higher frequencies, and the spectral intensity becomes a U-shape. This can be attributed to the temporal walk-off between the interacting pulses. Inspecting the temporal overlap of IPDFG and pump pulses indicates the complete temporal separation between them due to their different group velocity in the crystal, as shown in figure \ref{fig:fig3}-b, figure \ref{fig:fig3}-c, and figure \ref{fig:fig3}-d. The input pump pulses have a temporal pedestal, which originates from the residual higher-order phase on the pump pulses due to the dispersive mirrors and soliton self-compression in the fiber. The residual higher-order phase caused by the dispersive mirrors is mainly at the two limits of the input pump spectrum, leading to the generation of high-frequency wings of the IPDFG spectrum beyond the temporal walk-off of the interacting pulses (see figure \ref{fig:fig3}-d). Furthermore, soliton self-compression causes the center of the pump spectrum to carry higher-order phases. These phases manifest as a temporal pedestal, which in turn leads to the generation of low-frequency wings in the IPDFG spectrum. Consequently, for crystal thicknesses greater than 0.5\,mm, the spectrum becomes U-shaped.

\begin{figure}[h]
\begin{center}
  \includegraphics[width=0.8\linewidth]{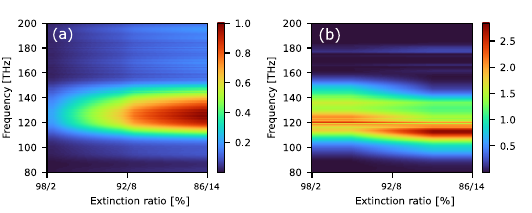}
  \caption{\textbf{Spectral evolution with respect to waveplate angles.} (a) Numerically simulated spectral evolution and (b) experimentally measured spectral evolution of the IPDFG with respect to the waveplate angle. The measured spectra are obtained from the Fourier transform of the corresponding EOS trace.} 
  \label{fig:fig4}
  \end{center}
\end{figure}

The energy distribution among the orthogonal components of pump pulses is crucial for maximizing the efficiency of difference frequency generation in type I phase-matching. This optimization relies on the principles of energy and momentum conservation. During the down-conversion process, the energy of the high-frequency components of the o-pump decreases, as illustrated in figure \ref{fig:fig3}-e. Conversely, the low-frequency components of the e-pump are amplified (figure \ref{fig:fig3}-f). Consequently, a series of simulations have been conducted to investigate the impact of the polarization distribution of pump energy on the efficiency of the down-conversion process in a 0.5\,mm-thick type I BiBO crystal. The energy of the e-pump pulses was scaled from 2\% to 22\%, while the total energy of the pump pulses remained constant. Figure \ref{fig:fig4}-a shows the generated IPDFG spectra at different pump polarization ratios. This assumes the placement of an ideal, dispersion-free, broadband waveplate before the BiBO crystal. The simulation indicates that the conversion efficiency peaks at a e-pump ratio of 22\% and decreases for higher pump polarization ratios.

\subsection*{Experimental results}

As discussed above, employing waveplates or rotating the crystal axes in the laboratory frame imposes constraints for polarization-sensitive measurement. Therefore, we took advantage of the fiber symmetry for ideal and dispersion-free tunning of the polarization state of the octave-spanning pump spectrum. A half waveplate (Altechna 2-APW-L2-018C) was placed at the relatively narrowband input of the second gas-filled HCF. A 6-inch focal length parabolic mirror was employed for focusing the pump to a beam size of 44 $\mu$m (FWHM), resulting in the peak intensity of 300\,$TW/cm^2$. A 0.5\,mm-thick type I, BiBO crystal cut in the XZ plane and at a phase-matching angle of $\theta$ = 11\,$^o$ was used for IPDFG. The crystal was positioned behind the focus to mitigate damage. Afterward, the IPDFG beam was collimated to a $1/e^2$ beam diameter of 3.2\,mm utilizing a 4-inch focal length parabolic mirror. A custom-designed broadband dichroic beam splitter (UFI BS2214-RC2) separated the pump and IPDFG beam. A custom-built double-angle chirped mirror compressor (UFI IR7202) with four reflections was used to compensate for the accumulated dispersion on the IPDFG pulses due to propagation in the BiBO crystal, refractive optics and air, yielding 15\,fs (FWHM) CEP-stable pulses. 

The half waveplate at the input of the second gas-filled HCF was used to tune the polarization extinction ratio of the pump pulses at the IPDFG szage. Starting at the system's overall extinction ratio of 98\%- 2\% (o-e), the half waveplate angle was tuned, and the corresponding IPDFG output power was measured. The average power was measured by placing a thermal power meter after a holographic wire grid polarizer angled to allow linearly polarized light with no rotation in the laboratory frame. The power of IPDFG pulses was 85\,mW at an extinction ratio of 98\%-2\% (o-e), with a maximum power of 185\,mW at an extinction ratio of 86\%-14\%. The electric field of the IPDFG pulses was characterized via electro-optic sampling (EOS) comprising a 20\,$\mu$m BBO crystal \cite{srivastava2023near, Keiber2016-xv}. The retrieved spectra from the measured electric field of the IPDFG pulses at various extinction ratios are shown in Figure \ref{fig:fig4}-b. IPDFG conversion efficiency gradually increases until the extinction ratio is 86\%-14\% (o-e). The minor discrepancy between the measured and simulated optimum polarization ratio could be associated with the slight birefringence in the fiber due to fabrication imperfections. The birefringence of the fiber could lead to uneven phase retardation resulting in elliptical polarization.

\begin{figure}[h]
\begin{center}
  \includegraphics[width=0.8\linewidth]{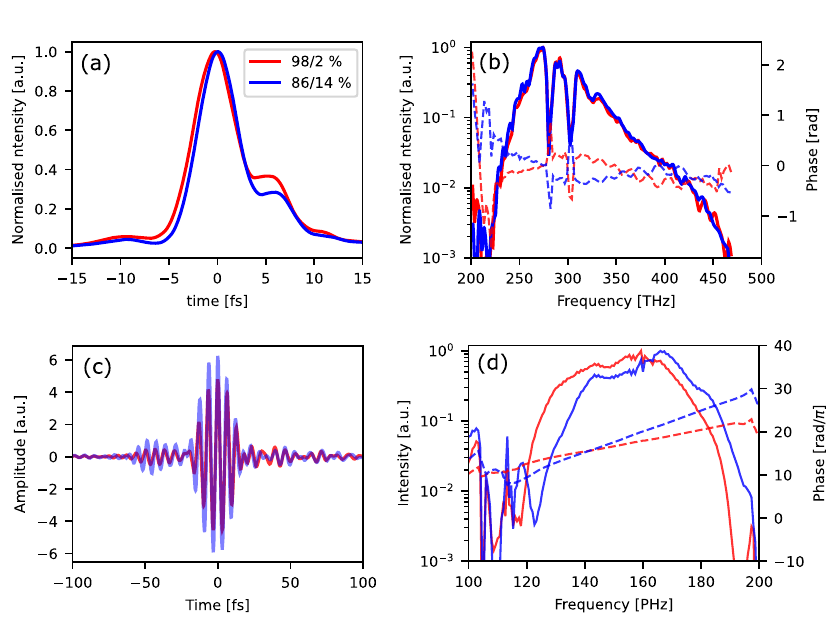}
  \caption{\textbf{Polarization optimisation.} The retrieved (a) temporal profiles and (b) retrieved spectra of pump FROG characterization for two waveplate angles. (c) The characterized electric field of the IPDFG pulses via EOS at two extinction ratios and their corresponding spectral intensity are shown in (d). The dashed line indicates the phase of the spectrum in both (b) and (d). The legend in (a) applies to all figures.}
  \label{fig:fig5}
  \end{center}
\end{figure}

The temporal profile of the output pulses from the fiber and the generated IPDFG pulses at two extinction ratios were analyzed to confirm that the polarization tuning of the input pulses does not affect the dynamics of soliton self-compression. For this purpose, SH-FROG was employed to characterize the fiber's output pulses. Furthermore, to completely characterize the IPDFG pulses, the electric field was measured through EOS. Figure \ref{fig:fig5}-a and figure \ref{fig:fig5}-b show the retrieved temporal profiles and retrieved spectra of the octave-spanning output pulses from the fiber at two polarization ratios of 98\%-2\% (o-e) and 86\%-14\% (o-e). The measurements indicate that polarization tuning has a negligible impact on the envelope and spectrum of the fiber's output. This suggests that the soliton dynamic inside the fiber remains unaffected due to the minimum birefringence of the fiber. The corresponding electric field of the generated IPDFG pulses at the two polarization ratios are shown in figure \ref{fig:fig5}-c. It has been observed that while polarization tuning enhances the amplitude of the IPDFG waveform, the electric field of the generated pulses remains unchanged.

\section{\label{sec:level5}Conclusion}
Over recent decades, the generation of high-energy, high-power pulses has increasingly relied on spectral broadening in fibers or multi-pass compression. IPDFG stands out as a highly promising method for generating octave-spanning pulses in the SWIR and MIR spectral regions, with unparalleled CEP stability from these frontends. Type I phase-matching, in particular, is favored for its superior bandwidth and conversion efficiency, albeit it necessitates an optimized polarization distribution along the crystal axes.

In this work, we conducted a detailed numerical analysis to identify the optimal polarization distribution within type I BiBO crystals for the generation of octave-spanning, CEP-stable pulses in the SWIR region. Prior approaches primarily utilized broadband waveplates or involved altering the orientation of the crystal axes with respect to the laboratory frame for optimizing the down-conversion efficiency. To circumvent the limitations presented by these methods, we have shown that in ultrashort pulse generation from gas-filled hollow-core photonic crystal fibers, the efficiency of IPDFG can be maximized by adjusting the polarization ratio of the narrowband pulses prior to spectral broadening. By implementing a narrowband waveplate before soliton compression, we successfully tuned the polarization of octave-spanning, near-single-cycle pulses for efficient IPDFG without affecting the soliton dynamics within the fiber. Our findings indicate that the orientation of polarization relative to the fiber does not impact the duration or the spectrum of the soliton self-compressed pulses. Additionally, we rigorously validated our results by directly characterizing the electric field of the IPDFG pulses, noting that while the polarization rotation of the input pulses enhances the IPDFG pulse amplitude, their spectral phase remains well-behaved.

The approach discussed in this work is not only applicable to HCF systems but can be extended to other symmetrical systems like multi-pass geometries, or multi-stage spectral broadening \cite{nisoli2024hollow, Fattahi:16, Viotti:22, Lu:14, Barbiero:21}. These results are particularly vital for broadband down-conversion of high-power systems to overtone and fingerprint spectral regions, where the bandwidth and dispersion characteristics of waveplates present challenges, and in experiments demanding pure polarization in the eigenmodes of the laboratory frame.

\begin{acknowledgments}
We thank Gunnar Arisholm for his support in the numerical analysis presented in this work.
\end{acknowledgments}

\section*{DECLARATIONS}
\begin{itemize}
    \item This work was supported by research funding from the Max Planck Society.
    \item Conflict of interest/Competing interests: The authors do not declare any competing interests.
    \item Authors' contribution: A.S. and K.S. conducted the measurements. A.S. and A.H. developed the frontend. A.S. and S.J. performed the simulations. A.S. and H.F performed the data analysis and wrote the manuscript. All authors proofread the manuscript.
\end{itemize}

\section*{DATA AVAILABILITY}
The data that support the findings of this study are available from the corresponding author upon reasonable request. 
\bibliography{aipsamp}

\end{document}